\documentclass[11pt]{article}
\usepackage[dvips]{graphicx}
\def\L{{\cal L}}
\begin{document}
\begin{center}
\bf{
NUCLEAR MATTER MEAN FIELD WITH EXTENDED NJL MODEL} \\ \vspace{0.5cm}
Steven A.\ Moszkowski\\
{\sl \small UCLA, Los Angeles, CA 90095, USA} \\\vspace{0.5cm}
Constan\c{c}a Provid\^encia, Jo\~ao da Provid\^encia and  Jo\~ao M. Moreira \\
{\sl \small University of Coimbra, P-3000 Coimbra, Portugal} \\ \vspace{0.5cm}
\rm \today
\end{center}

\begin{abstract}
An extended version of the Nambu-Jona-Lasinio (NJL) model is
applied to describe both nuclear matter and surface properties of
finite nuclei. Several parameter sets  are discussed and a
comparison of the saturation properties and equation of state
(EOS) with the NL3 parametrization of the non-linear Walecka model
is made. The properties of asymmetric matter are discussed.
\end{abstract}


\def\d{{\rm d}}
\def\L{{\Lambda}}
\section{Introduction}

Strong interaction dynamics of mesons and baryons is believed to be described
by QCD. This theory exhibits a non perturbative behaviour at low energies,
a circumstance which renders the analytic study of the theory rather difficult.
The NJL model is a popular substitute which has in common with QCD important
symmetries of the quark-flavour dynamics. This model has been very successful
in the description of the meson sector. The question arises whether the model
also allows for inhomogeneous solutions at non-zero density which are of interest
for the description of hadronic matter.

The NJL model \cite{Nambu61} was originally developed for the
purpose of understanding hadron physics. In this model, hadron
masses are generated by spontaneous symmetry breaking of the
vacuum. In the modern form of the NJL model, we start with
essentially massless quarks interacting via zero range
interactions, but with a cutoff in momentum space
\cite{Vogl91,Klevansky92,Hatsuda}. NJL is thus only an
effective theory, in which form factors and other finite range
effects have been ignored, and it does not take into account quark
confinement. However, it may work quite well in the region of
interest in nuclear physics, i.e. for excitations less than the
scalar meson mass.

The paper has been organized as follows: It is shown that  with
the generalization of the NJL model  for nucleons,  referred
to as the Extended NJL (ENJL) model, it is possible to get a
reasonable nuclear equation of state and behaviour of the
effective mass.  Then we list some numerical results, concerning
both nuclear matter and quark matter. It is found that the
performance of the ENJL model is almost perfect as far as bulk
static properties, such as saturation and binding energy of
nuclear matter, or the effective mass of nucleons,  are concerned.
However, the model fails when we wish to deal with dynamical
properties, since it leads to scalar excitations with the
unacceptable mass of twice the nucleon mass, or to a poor
description of the nuclear surface. In order to overcome these
drawbacks, the ENJL model is replaced by a
chiral invariant model of nucleons interacting through the
exchange of $\sigma\,,\vec\pi$ and $\omega$ fields, which we call
extended chiral sigma (ECS) model. For the description of bulk
static properties, the ECS model is almost equivalent to the ENJL model.
In these models, the nucleons are regarded as composite particles.
The composite nature of the particles reflects itself in the fact
that the coupling depends on the local density. The surface
properties of finite nuclei are calculated within the ECS model.
Some remarks concerning the connection of the ENJL and ECS models
to other relativistic chiral models \cite{boguta83,furn93} are
presented. Then, some properties of asymmetric nuclear matter are
discussed and some brief conclusions are drawn. Next, we discuss a
two dimensional version of the NJL model for which some simple
analytic results can be obtained. This is followed by a comparison
of our model with several other models, including the quark meson
coupling model and versions of relativistic mean field models.
Preliminary results of the present work have been presented in
\cite{proc}.

\section {NJL model}
\subsection{Original NJL Model}
The NJL model \cite{Klevansky92} is defined by the Lagrangian density
\begin{equation}{\cal L}=\bar\psi(i\gamma^\mu\partial_\mu)\psi+G[(\bar\psi\psi)^2
+(\bar\psi i\gamma_5\vec\tau\psi)^2].\label{NJL1}
\end{equation}
A regularizing momentum cut-off $\Lambda$ is part of the model.
The Lagrangian is equivalent to the Hamiltonian
\begin{equation}
{\cal H}_{NJL}=\sum_{k=1}^N\vec{p}_k\cdot\vec{\alpha}_k+
G\,\sum_{k,l=1}^N\delta(\vec{r}_k-\vec{r}_l)
\beta_k\beta_l(1-\gamma^5_k\gamma^5_l\vec{\tau}_k\cdot\vec{\tau}_l)\,.
\end{equation}
The vacuum is described by a Slater Determinant $|\Phi_0\rangle$
constructed from plane waves which are negative energy eigenfunctions of the single particle Hamiltonian
$h=\vec{p}\cdot\vec{\alpha}+\beta m.$ The ``constituent
mass'' $m$ is a variational parameter.

If moreover positive energy eigenfunctions with momentum
$\vec p$ satisfying $|\vec p|<p_F$ are occupied, so that $p_F$ is the
Fermi momentum, the energy expectation value
$E=\langle \Phi_0|{\cal H}_{NJL}|\Phi_0\rangle$ is given by
\begin{equation}\label{NJL0}
E = - \nu\, \sum_{p_F\leq|\vec p|\leq\Lambda}\frac{p^2}{\sqrt{m^2+p^2}}
-{m^2\,G}\,\nu^2\,[\sum_{p_F\leq|\vec p|\leq\Lambda}
\frac{1}{\sqrt{m^2+p^2}}]^2\,.\label{EnerNJL}
\end{equation}

For quark matter, the degeneracy is $\nu=2N_cN_f$ and $\L$ is such
that $m=313$ MeV is the constituent quark mass in the vacuum.
The gap equation, which determines $m$, reads,
\begin{equation}2\,G\,\nu \sum_{p_F\leq|\vec p|\leq\Lambda}
\frac{1}{\sqrt{m^2+p^2}}=1\,. \label{GapEq}
\end{equation}

\subsection{Nuclear Matter vs Quark Matter Cutoff}
It is well known that the NJL model does not contain a mechanism for quark
confinement. Thus the clustering of quarks into nucleons is not accounted
for in this model. In this paper, quark clustering is put in by hand. The
way it is done here is that we treat the nucleons as the elementary particles,
but take quark structure into account in the interactions.
We will find it convenient to define dimensionless integrals $I_n$ by:
\begin{equation}
\int_{0}^{x}\frac{2p^2}{(1+p^2)^{n-\frac{1}{2}}} dp = I_n(x) \,.\\
\end{equation}
According to the NJL model, the volume integral of the $NN$
interaction is given by
\begin{equation}\label{Vs}
\widetilde{V}_{s} = -4\frac{\pi^2}{I_2(\frac{\Lambda \,N_c}{M_0})}
N_c\,\mu\,c^2 (\frac{\hbar}{\mu\,c})^3 = -8.99\, N_c^3 M_0\,c^2
(\frac{\hbar}{M_0\,c})^3 \,.
\end{equation}
Here $N_c$ is the number of colors, and $M_0$ is the free nucleon
mass. For quark matter, $N_c=3$. The scalar $(\sigma)$ meson mass is:
\begin{equation}
\mu = m_s = \frac{2M_0}{N_c} \,.
\end{equation}
Let us now try to get the same $NN$ volume integral with a
single color. This can be accomplished by using a lower value of
the cutoff, denoted here by $\Lambda \,'$. Further, from here on,
we will often express all masses in units of the free
nucleon mass $M_0$. Also, we will set $\hbar$ and c equal to 1. We find that: \\
\begin{equation}\label{21}
I_2(\Lambda \,') =  \frac{I_2(\Lambda \, \,N_c)}{N_c^3}   \,.\label{I2Lambda'}
\end{equation}
 An alternative way of taking quark clustering into account is by fixing $\Lambda\,'$
so that the free nucleon mass in vacuum, $M_0$ is preserved.
Actually, such a prescription has been adopted in some of the calculations
which are reported in Section 4 of this paper. However, the
value of $\Lambda\,'$ so obtained is in a very reasonable
qualitative agreement with the one determined by eq.
(\ref{I2Lambda'}).
\subsection{Extended NJL Model}
 The NJL model can be extended \cite{Koch87,
Bentz01,proc,Msg04}
 to yield reasonable saturation properties of nuclear matter, the field
$\psi$ being then the nucleon field. In Ref. \cite{Bentz01}, the nucleon is 
constructed as a 3-quark bound state, this
quark substructure giving rise to a mechanism for saturation of nuclear
matter which plays the same role as the 8-fermion term in our ENJL model.
Such substructure may therefore be regarded to provide a microscopic basis
for our phenomenological approach. An effective density
dependent coupling constant is obtained if the following extended
NJL (ENJL) Lagrangian density, which actually pushes chiral symmetry
restoration to higher densities, is considered,
\begin{eqnarray}
{\cal L}&=&\bar\psi(i\gamma^\mu\partial_\mu)\psi+G_s[(\bar\psi\psi)^2
+(\bar\psi i\gamma_5\vec\tau\psi)^2]-G_v(\bar\psi\gamma^\mu\psi)^2\nonumber\\
&-&G_{sv}[(\bar\psi\psi)^2+(\bar\psi i\gamma_5\vec\tau\psi)^2]
(\bar\psi\gamma^\mu\psi)^2.\label{lagran}
\end{eqnarray}
For nuclear matter $\psi$ is the nucleon field, the degeneracy is
$\nu=2N_f$ and  $\L$ is such that $M=939$ MeV is the nucleon mass
in the vacuum, determined variationally, as explained below.
A different extension which also leads to nuclear saturation has been made 
\cite{Bur03,Bohr04},
This involves the fourth power of the scalar density.
Another alternative way of extending the NJL model was studied by Rezaeian and Pirner \cite{Res05}. 
Here the nucleon is considered to be a composite particle, a bound state of a quark and a diquark. 
The term in $G_v$ is supposed to simulate a chiral
invariant short range repulsion between nucleons
connected, possibly, with the chromomagnetic interaction between quarks.
The term in $G_{sv}$ accounts for the density dependence of the
scalar coupling between nucleons which is a manifestation
of the composite nature of these particles
and is equivalent to  allowing  the coupling constant $G$ in the original NJL model to depend on $\rho$
in such a way that
$G$ becomes replaced by $ G_s(\rho) =
{G_s}(1-\frac{G_{sv}}{G_s} \rho^2)
$.
It is found that the magnitude of the $\sigma$ field
decreases as the hadronic density increases,
in accordance with the corresponding tendency for chiral symmetry restoration.
This mechanism has a similar effect, for saturation, as the analogous one
proposed by Guichon \cite{Guichon},
according to which confinement is
affected by an increase of the density in such a way as to become
gradually less effective, which amounts to an increasing repulsion.
For nuclear matter, the NJL model leads to binding but
the binding energy per particle does not have a minimum except
at a rather high density where the nucleon mass is small or vanishing. Thus the pure NJL
model does not give a proper equation of state. The introduction of  the  $G_{sv}$ coupling term
is required to correct this. On the other hand, for
quark matter we set $G_v=G_{sv}=0$. For quark matter
the energy per particle $W$ predicted by NJL does not lead to binding and does
not have a minimum except for unreasonably small
values of the cutoff momentum $\Lambda$.

\begin{table}
\begin{tabular}{cc|cc}
\hline
&  &
\multicolumn{1}{c} {\bf Coupling constants, etc.}\\
{ \bf Quantity} & { \bf  EOSI} &{ \bf Quantity} & { \bf  EOSI} \\
$G_s$(fm$^2$)     &    3.880&$M_0$ (MeV)    &     939 \\
$G_v$(fm$^2$)           & 3.952&$\Lambda$ (MeV)      &418.9\\
$G_{sv}$(fm$^8$)        &  -4.901&$I_1$&0.0560 \\
$G_\rho$(fm$^2$)           & 2.794& $I_2$&0.0502\\
$c_v$  & 0.913&$\rho/\rho_0$ ($M=0$)&  3.0  \\
$c_{sv}$  & 0.427& $E_{Bq}=E_{BN}$ (MeV)& 111   \\
$\tilde{V_S}$  & -1705.4 & $\rho/\rho_0$ ($E_{Bq}=E_{BN}$)
&4.3\\
\hline
\end{tabular}
\caption{The coupling constants and properties for our preferred equation of state}
\label{tab:exp}
\end{table}

The  thermodynamical potential per volume corresponding to (\ref{lagran}) is
\begin{equation}
\omega(\mu)=\langle\bar\psi(\vec\gamma\cdot\vec p)\psi\rangle
-G_s\langle\bar\psi\psi\rangle^2+G_v\langle\psi^{\dag}\psi\rangle^2
+G_{sv}\langle\bar\psi\psi\rangle^2\langle\psi^{\dag}\psi\rangle^2-\mu
\langle\psi^{\dag}\psi\rangle\label{omega}
\end{equation}
where exchange terms have been neglected. By
$\langle\bar\psi\Gamma\psi\rangle$ we denote the 
expectation value per volume.
$\langle\bar\psi\Gamma\psi\rangle=
\frac{1}{V}\langle\Phi_0|\sum_k\beta_k\Gamma_k|\Phi_0\rangle\nonumber$. The condition
$\partial\omega/\partial M=0$ implies
$$M=-2G_s\langle\bar\psi\psi\rangle+
2G_{sv}\langle\bar\psi\psi\rangle\langle\psi^{\dag}\psi\rangle^2.
$$
Here $M$ is the $\mu$ dependent
                nucleon mass. To avoid a cumbersome notation, the $\mu$
                dependency is not explicitly indicated but is implicitly
                assumed, without danger of confusion. The free nucleon
                mass $M_0$ is the value of $M$ for $\mu=0$. 
The condition $\partial\omega/\partial p_F=0$ implies
$$E_{p_{F}}=\mu-2G_v\langle\psi^{\dag}\psi\rangle-
2G_{sv}\langle\psi^{\dag}\psi\rangle
\langle\bar\psi\psi\rangle^2,
$$
with $E_{p_F}=\sqrt{M^2+p_F^2}$.
These conditions fix the values of $p_F,\,M$ for given $\mu.$ We observe that the gap
equation, which determines $M,$ is the same as eq. (\ref{GapEq})
except for the replacement of $ G$ with
${G_s}(1-\frac{G_{sv}}{G_s} \rho^2).$
 The choice with $G_s$ as a phenomenological
parameter fixed so that nuclear matter saturation properties are
reproduced, we will refer to as model I. The value of $G_s$ is about 20\% 
smaller than that obtained from the NJL model. On the other hand, it is
also tempting to choose
 $G_s$ for nuclear matter
nine times bigger than the quark matter value, $G$. The philosophy
behind the last assumption is that the NN interaction is due to
the instanton interaction between quarks, predicted by QCD in the
weak coupling regime. This choice of $G_s$ defines what we will
denote in the sequel as model II.

The properties of the extended NJL model are now easily computed.
\begin{table}
\begin{tabular}{lcccc}
\hline
      &{\bf EOSI} &  {\bf NL3}\\
\hline
{ $E/A-M_0$(MeV)} & {-16.12}  & { -16.3}\\
$\rho_0$ (fm$^{-3})$   &{   0.148} &0.148 \\
{ $M/M_0$} &{ 0.75} &   0.60\\
$K$ (MeV) &{  295}   &272 \\
{ $W_{surf}$ (MeV)}&{ 19.43}  &19.38\\
$R$ fm&5.33  &5.328\\
$t$ fm &{ 2.64}   &2.65\\
\hline
\end{tabular}
\caption{Nuclear matter saturation properties and surface properties}
\label{tab:sur}
\end{table}

\section{Properties of nuclear matter}
In Table \ref{tab:exp}, numerical results pertaining to our
preferred model EOSI are presented. As input, we use the
saturation properties of nuclear matter fitted empirically in
Ref.\cite{furn}.
\begin{figure}
\includegraphics[width=.6\textwidth,angle=0]{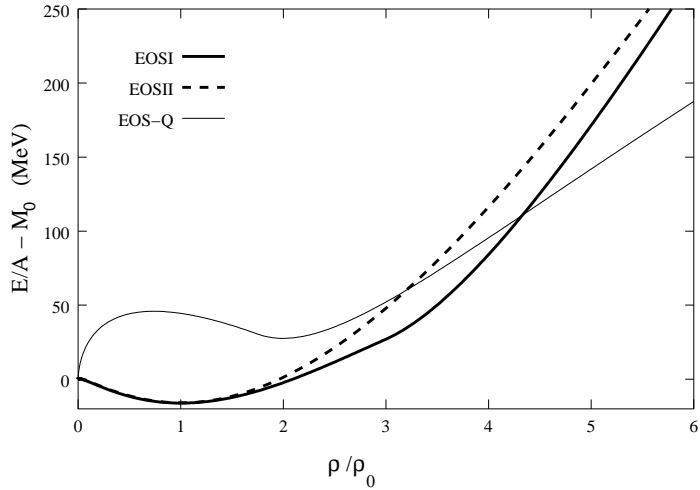}
\caption{Energy vs density in ENJL Model for symmetric nuclear
matter (EOSI and EOSII) and for
  quark matter (EOSI-Q)}
\label{fig1}
\end{figure}
The three coupling constants $G_s, G_v$ and $G_{sv}$ are adjusted
to fit an effective mass $M(\rho_0) = 0.75 M_0$ and the saturation energy
and density.
Model EOSI seems to be quite realistic, since it fits well both
saturation properties and the nuclear surface properties. We also
compare our results with those of
\cite{ring} in which a relativistic  mean field theory (RMF) is
used, and which describes well  both stable and unstable nuclei.
The RMF parametrization introduced in \cite{ring} is known as NL3.
Model I gives results  quite close  to NL3 especially for the
surface properties (see Table \ref{tab:sur}). The relation between
the coupling constants $G_s,\,G_v,\, G_{sv}$ and the parameters
$c_v,\,c_{sv}$, which will be used later, is fixed as follows:
\begin{eqnarray}
G_s &=& \frac{\pi^2}{2 I_1(\Lambda')\,M_0^2} \\
G_v &=& \frac{1}{2}c_v \frac{\widetilde{V_s}}{\hbar\,c}=\frac{\pi^2\,c_v}{2 I_2(\Lambda')\,M_0^2} = G_s\,\frac{I_1(\Lambda')}{I_2(\Lambda')}\,c_v  \\
G_{sv}&=& -\frac{\pi^6\,c_{sv}}{2\,I_1(\Lambda')^2\,I_2(\Lambda')\,M_0^8} = -G_s\,\frac{\pi^4\,c_{sv}}{I_1(\Lambda')\,I_2(\Lambda')\,M_0^6}\,.
\end{eqnarray}
Here $M_0 = 4.76 fm^{-1}$. 
The crucial role played by clustering must be stressed. It should
be pointed out that without clustering there is no real binding
and that, moreover, the incompressibility becomes unacceptable. In
Fig. \ref{fig1} we plot the binding energy for nuclear matter and
quark matter (EOS-Q). For the latter case, we use a quark NJL
model which differs from the nucleon NJL model (ENJL) used in the
rest of this paper. At some density, the nucleon matter curve
intercepts the quark matter curve, so it is clear that, at high
density, quark matter prevails and may be found, for instance, in
the core of neutron stars. 
Our experience with this problem tells us that without clustering and without the eight fermion interaction (that is, the density dependent coupling)
saturation comes closely after chiral symmetry restoration, almost coinciding with it. Clustering and the eight fermion interaction places chiral symmetry 
restoration well after saturation. \\ 
The philosophy underlying model EOS II is to use a coupling $G_s$ 9 times bigger than G in the quark NJL model and to fit one less quantity. Thus the effective mass at saturation was left free. We have taken $E/A=-15.8$ MeV
and $\rho_0=0.148$ fm$^{-3}$, i.e. $k_F = 0.130$ fm$^{-1}$
according to \cite{furn,ring}. Using this model, one gets quite
different results for the coupling constants from EOS I, namely $G_s$ = 1.746 fm$^2$, $G_v$ = 3.387 fm$^2$, $G_{sv}$ = -1.839 fm$^8$, $\Lambda$ = 553 MeV  . The effective mass
at the saturation density turns out to be 0.89, much closer to the
free nucleon mass than for model I. Also, this model leads to
surface energy and thickness only about half of the empirical
values. Similar results are obtained by Koch et al,
ref. \cite{Koch87}. However, the cutoff $\Lambda$  obtained using EOS II is not far from the one determined by eq. (\ref{21}).
\begin{figure}
\includegraphics[width=.6\textwidth,angle=0]{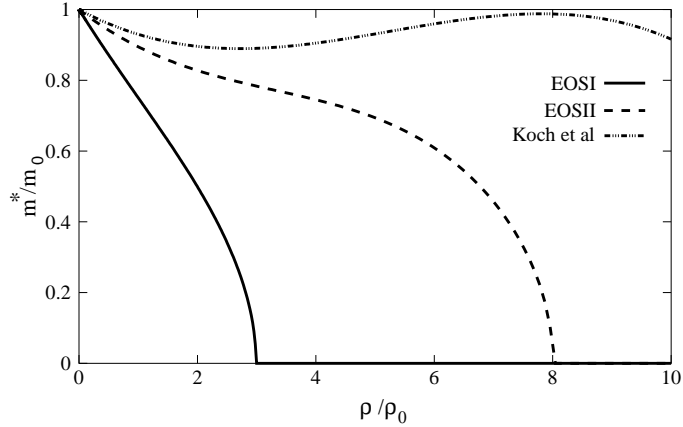}
\caption{Effective Mass vs density in ENJL Model for models I and II and results 
of Koch et al. \cite{Koch87}}
\label{fig2}
\end{figure}
\section{Surface properties}
Next we will calculate nuclear surface properties which are mainly
determined by the scalar meson, the meson responsible for the
nuclear attraction. We have seen that the ENJL model accounts
adequately for bulk static properties of hadronic matter, such as
saturation and binding energy of nuclear matter, or the effective
mass of nucleons. However, the model fails when we wish to deal
with dynamical properties, since, as it has been shown \cite{proc,
Bohr04}, it leads to scalar excitations with the unacceptable mass
of twice the nucleon mass, and to a poor description of the
nuclear surface.  We keep the ENJL Lagrangian, in spite of the 
referred drawbacks, because it is extremely simple and
transparent, clarifies the underlying physics and  incorporates 
in the simplest possible manner the basic philosophy of our approach. \\ 
In the extended chiral sigma model, introduced in \cite{proc,Bohr04},
the nucleons and mesons are regarded as composite particles. The version
used in \cite{Bohr04} is referred to as ESM. In the present paper and in \cite{proc}, a slightly different version is used which we refer to as ECS. Both models are equivalent to corresponding ENJL models, as far as ground state properties are concerned, provided exchange terms are neglected. However, there are significant differences between the models. In ESM, the coupling to the sigma meson depends on the scalar-pseudoscalar
field, while in the ECS model, that coupling depends on the baryon density.
Moreover, the nuclear calculations are more easily implemented in the ECS model, which is free of instability that may occur in ESM. The advantage of the ECS model over ENJL stems from the fact that it generates adequate collective properties, this being the reason that ECS can be used to study the nuclear surface.  \\  
We write formally  the ECS model Lagrangian density  as  
\begin{eqnarray}
{\cal L}&=&\bar\psi(i\gamma^\mu\partial_\mu)\psi
-g_s\,(\sigma\,\bar\psi\psi+\bar\psi
i\gamma_5\vec\pi\cdot\vec\tau\,\psi)
\left[1+a_1(\bar\psi\gamma_\mu\psi)(\bar\psi\gamma^\mu\psi)\right]^{1/2}\nonumber
- g_v \, \bar\psi V^\mu\gamma_\mu\psi\,\\&+&{1\over2}(\partial^\mu\sigma\partial_\mu\sigma+
\partial^\mu\vec\pi\cdot\partial_\mu\vec\pi)
-{1\over2}m_s^2(\sigma^2+\vec\pi\cdot\vec\pi)-{1\over4}V^{\mu\nu}V_{\mu\nu} + \frac{1}{2}
m^2_v\, V^\mu V_\mu\,,\label{lagr2}
\end{eqnarray}
with
\begin{eqnarray}
\frac{g_s^2}{2\, m_s^2}=G_s, \qquad \frac{g_v^2}{2\, m_v^2}=G_v,
\qquad \frac{a_1\, g_s^2}{2\, m_s^2}
=-G_{sv}.
\end{eqnarray}
For comparison, the ESM Lagrangian density is given by Eq.(11) of \cite{Bohr04},
while in the linear sigma model, ref. \cite{Lee81}, we
have:
\begin{eqnarray}
{\cal L}&=&\bar\psi(i\gamma^\mu\partial_\mu)\psi
-g_s\,\sigma\,\bar\psi\psi + {1\over2}
(\partial^\mu\sigma\partial_\mu\sigma+
\partial^\mu\vec\pi\cdot\partial_\mu\vec\pi)\nonumber \\
&-& \frac{m_s^2}{8 F_{\pi}^2}
(\sigma^2+\vec\pi\cdot\vec\pi-F_{\pi}^2)^2 \label{lagr3}
\end{eqnarray}
Here the nucleon mass is generated by a ``Mexican Hat" potential.
If we write $\sigma = \sigma_0 + F_{\pi}$ then the last term in
the linear sigma model contributes $-{1\over2}m_s^2 \sigma_0^2 ( 1
+ \frac{\sigma_0}{2 F_{\pi}})^2$ to the Lagrangian. 
It was shown in \cite{Fiolhais97} that  the Dirac-sea in  the
linear sigma model with valence and Dirac-sea quarks  but  no
``Mexican hat'' provides the above effective mesonic
self-interaction. \\
The quantity $F_{\pi}$ is equal to $\sigma=M/g_s$, when the energy is
minimized w.r.t.   $\sigma$. For zero particle density, we obtain,
according to the ECS model, Eq. (\ref{lagr2}),
\bf{$$\frac{g_s^2}{2m_s^2}=\frac{9\,g_s^2}{8\,M_0^2}=G_s=\frac{\pi^2}{2I_1(\frac{\Lambda}{M_0})M_0^2},
$$}\rm 
which leads to
\begin{equation}
F_{\pi}(ECS)=\frac{3M_0}{2\,\pi}\,\left[I_1(\frac{\Lambda}{M_0})\right]^{1/2}.
\end{equation}
This is to be compared to the pion decay constant according to the
ENJL model, which, for vacuum properties such as the present one, 
coincides with the nucleon NJL model,
\begin{equation}
F_{\pi}(NJL)  = \frac{M_0}{2\,\pi}\,\left[I_2(\frac{\Lambda}{M_0})\right]^{1/2}.\label{(20)}
\end{equation}
We observe that Eq. (\ref{(20)}) may be obtained from Eq. (3.26)(a) of Ref. \cite{Klevansky92} 
upon replacing the number of colours $N_c$ by 1 and $m^*$ by $M_0$.
The quantities $I_1$ and $I_2$ are defined in Section 2.2. The calculated pion decay constants for 
the ECS model (106.1 MeV) is in fairly
good agreement with the empirical value of 93 MeV, but 
the pion decay constants for the ENJL model (33.49 MeV) is not. 
This was expected since the performance of the ENJL model
is poor for dynamical properties. \\
 In order to describe surface properties we must
 include gradient
terms in the meson field
 \cite{pww99}. {  Somewhat analogously to our approach, in Ref. \cite{Bur03} supplementary gradient terms were introduced into some kind of extended NJL model with nucleons}.  
This is in consonance with the work \cite{Reinhardt87} where an effective field theory
of many-nucleon systems is obtained from the NJL type Lagrangian.
In the {ECS Lagrangian (\ref{lagr2})}, we will
keep only the gradient terms corresponding to the scalar meson,
the philosophy behind this approximation being that the relevant length scale is determined by the
sigma mass.

The corresponding thermodynamical potential per volume in the Thomas-Fermi approximation is then
\begin{eqnarray}
\omega(\mu)&=&\langle\bar\psi(\vec\gamma\cdot\vec p)\psi\rangle
+g_s\sigma\langle\bar\psi\psi\rangle
{ \left(1-{G_{sv}\over G_s}\langle\psi^{\dag}\psi\rangle^2\right)}^{1/2}\\
&+& g_v \,V_0\langle \psi^{\dag}
\psi\rangle+{1\over2}\vec\nabla\sigma\cdot\vec\nabla\sigma
+{1\over 2}m_s^2\sigma^2\, - \frac{1}{2}m_v^2\,
V_0^2\,-\mu\langle\psi^{\dag}\psi\rangle \nonumber
\end{eqnarray}

Minimization with respect to $\sigma,\,V_0$ yields eq.
(\ref{omega}) for infinite matter, provided we choose $G_s=g_s^2/(2
m_s^2)$.
\begin{figure}
\includegraphics[width=.6\textwidth,angle=0]{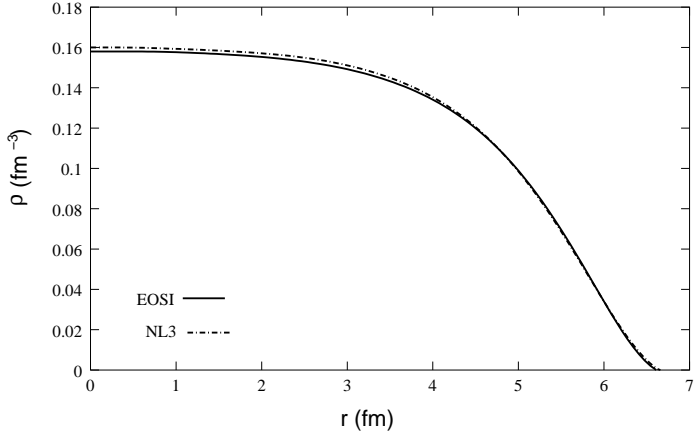}
\caption{Density profile for nuclei with  A = 100}
\label{fig3}
\end{figure}
Variation with respect to $M$, $p_F$,  $\sigma$ and $V_0$ gives:
\begin{eqnarray}
&&M=g_s\sigma{{\left(1-{G_{sv}\over G_s}\langle\psi^{\dag}\psi\rangle^2\right)}}^{1/2}\,,\nonumber\\
&&E_{p_{F}}=\mu-g_v\, V_0
-g_s\sigma{G_{sv}\over G_s}\langle\bar\psi\psi\rangle\langle\psi^{\dag}\psi\rangle{\left(1-{G_{sv}\over G_s}\langle\psi^{\dag}\psi\rangle^2\right)}^{-1/2}\,,\nonumber\\
&&\nabla^2\sigma=m_s^2\sigma-g_s\langle\bar\psi\psi\rangle
{\left(1-{G_{sv}\over G_s}\langle\psi^{\dag}\psi\rangle^2\right)}^{1/2}\,,\nonumber\\
&&0=m_vV_0-g_v\langle\psi^{\dag}\psi\rangle\,.
\end{eqnarray}
Next, we look for a droplet solution within model I and compare
its surface properties with the results obtained with the NL3
parametrization \cite{ring}. In the small surface thickness
approximation ($\nabla^2 \sigma\sim \frac{d^2 \sigma}{dr^2}$) the
free energy $F$ can be rewritten as \cite{Nielsen90}:
\begin{equation}
F=\int 4\pi r^2 d r{\left( \left(\frac{d \sigma}{d r}\right)^2-C \right)}+\mu A,
\end{equation}
where $C$ is a constant and $A$ is the number of particles. For droplets with radius $R$ and volume $V$,
\begin{equation}
F(R)=W_{surf} \, A^{2/3} - CV + \mu \,A. \label{free}
\end{equation}
The surface energy per unit area of these droplets in the small thickness approximation is
\begin{equation}
W_{surf}=\frac{4\pi\, R^2}{ A^{2/3} }\int_0^\infty d r
\left(\frac{d \sigma}{d r}\right)^2. \label{sig}\end{equation} The
surface thickness $t$ is defined as the width of the region where
the density drops from $0.9 \rho_{B0}$ to $0.1 \rho_{B0}$, where
$\rho_{B0}$ is the baryonic density at $r=0$. In Table
\ref{tab:sur}, the values of $W_{surf},\, R,\, t$, for a nucleus
with $A=100$, predicted by  model I is displayed and compared with
the corresponding results obtained by the NL3 parametrization.
In the present approach the mass of the $\sigma$ meson appears as
an extra parameter. We have chosen $m_s=2\, m_q=626$ MeV, as in
the NJL model.
In Fig. \ref{fig3} we plot the density profile of a nucleus with
100 particles.

\section{Isospin asymmetric nuclear matter}

The recent advance in unstable nuclear-beam experiments has been providing us with new
knowledge on unstable nuclei away from the stability line. We  hope to get information on the
properties of dense matter under extreme conditions as the neutron-rich enviroment in neutron
stars.

\begin{figure}
\includegraphics[width=.6\textwidth,angle=0]{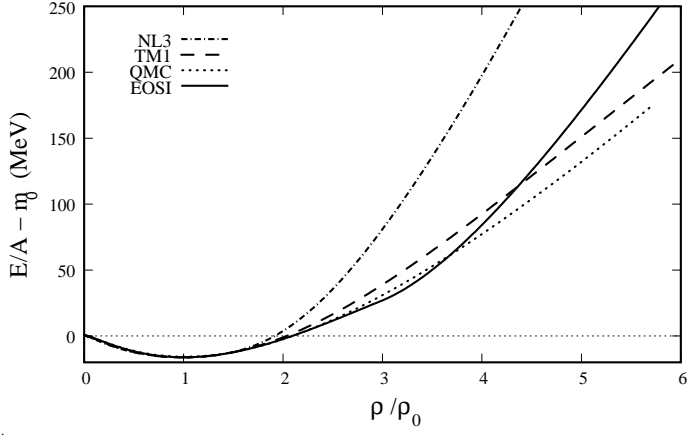}
\caption{Binding energy versus density according to QMC, EOSI, NL3
and TM1, for symmetric nuclear matter} \label{fig4}
\end{figure}

\begin{figure}
\includegraphics[width=.6\textwidth,angle=0]{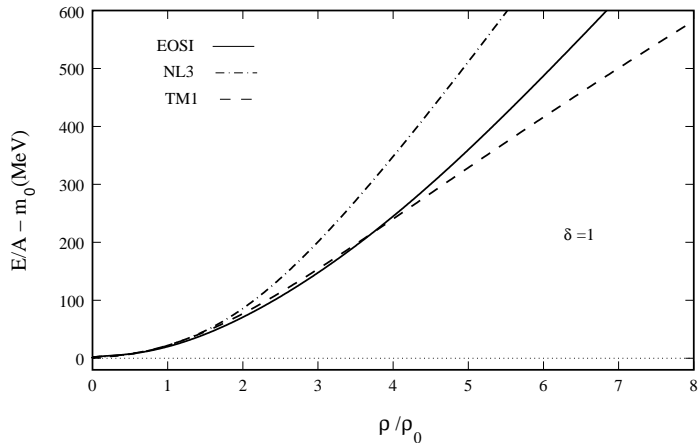}
\caption{Binding energy versus density according to EOSI, NL3 and
TM1, for neutron matter} \label{fig5}
\end{figure}

In order to describe isospin asymmetric nuclear matter within the
present formalism  it is important to include in the Lagrangian
density the isovector-vector term
$${\cal L}_\rho=-G_\rho \left[(\bar\psi\gamma^\mu \vec\tau\psi)^2+ (\bar\psi \gamma_5\gamma^\mu\vec\tau\psi)^2\right] .$$
Here, $G_\rho$ is chosen in such a way that the  experimental
value of the  symmetry energy {$a_{sym}=35$ MeV} is reproduced.\\
In Fig. \ref{fig4} we compare EOSI with two different
parametrizations of the non-linear Walecka
  model, NL3 \cite{ring} and TM1 \cite{toki95}, and the quark meson coupling model (QMC) \cite{Guichon}
  for symmetric matter. In QMC the structure of nucleons is taken
  into account having as underlying model the MIT Bag model.
  In Fig. \ref{fig5} we make a similar comparison with NL3 and TM1 for neutron
  matter. The RMF
  parametrizations were fitted to both stable and unstable nuclei. However TM1 contains a
  self-coupling term in the vector meson which weakens the contribution of the vector meson at
  high densities. We conclude that
EOSI is softer than NL3. Indeed, EOSI is also softer than TM1 and
QMC,
but becomes slightly stiffer
at higher densities, in connection with chiral symmetry
restoration.

The nuclear EOS, as function of density and asymmetry, is
{ reasonably well approximated by:
\begin{equation}
\frac{E}{A} = W =
|W_0|\,[(-2(1-\delta)\,\frac{\rho}{\rho_0}\,+\,(\frac{\rho}{\rho_0})^2]
\end{equation}}
where $\delta=(\rho_n-\rho_p)/(\rho_n+\rho_p)$. Here, $\rho_i,\,
i=n,p$ is the density of particles of type $i$.

\section{Original NJL Model in Two Dimensions}
For the two dimensional NJL model, we get simple analytic results.
The average kinetic energy $T$ is given by: \\
$T = \frac{1}{2}\,\rho_f - \frac{1}{6}\,\rho_f^2 + \ldots $, where
$\rho_f = 2\pi\,\rho_N$. Thus we get, ($c_{sv} = 0 $ for pure
NJL):
\begin{eqnarray}
m& =& 1 - \,g\,\rho_f - (g - \frac{1}{2})\,\rho_f^2 + \ldots \,,\\
W &=& -\frac{g\,-1}{2}\,\rho_f + 0\,\rho_f^2 + \ldots\,,
\end{eqnarray}
where $g = \frac{\widetilde{V_s}}{2\pi}$ and $m$ is the quark masss.
 For the two dimensional
case, we can obtain exact results for these two quantities, not
just for low densities:
\begin{eqnarray}
m^2& =& 1 - 2\,g\,\rho_f + (g - 1)^2 \,\rho_f^2  \,,\\
W &=& -\frac{g\,-1}{2}\,\rho_f \,.\label{82}
\end{eqnarray}
It is remarkable that higher order terms in $W$, cancel exactly. The vanishing term of order $\rho_f^2$
in Eq. (\ref{82}) has three components which cancel exactly:\\
1. Relativistic correction to energy momentum relation. \\
2. Increased kinetic energy due to reduced effective nucleon mass. \\
3. Increased attraction due to reduction in effective mass of scalar meson (similar to that of the nucleon).\\
For the more realistic three dimensional case, the physics is very similar, and again there is a near cancellation between these three terms. It cannot, of course, be perfect, since in three dimensions, these terms are proportional to $\rho^{4/3}$,\,$\rho^{5/3}$\, and $\rho^2$, respectively. This explains why
                  the $c_{sv}$ term is essential for achieving saturation.
\section{Connection with other Models}
\vspace{-0.25cm} 
\subsection{Guichon quark meson coupling model}
\vspace{-0.25cm}
One promising possibility is the quark-meson coupling model
proposed by Guichon \cite{Guichon}. In this model, the nucleon
mass is generated dynamically using a MIT bag model. Let us define
the dimensional coupling constant:
\begin{equation}
B_s = \frac{g_s^2}{m_s^2}\,\frac{\rho_0}{M} = \frac{\rho_0\,
\widetilde{V}_s}{M},
\end{equation}
where $\widetilde
                     V_s$ is the volume integral of the $NN$ interaction,
                     eq. (\ref{Vs}).
Then in the Guichon model, we automatically obtain a saturating term which can be put into our form:
\begin{equation}\label{saturating}
W_{\rho^2} = c_{sv}\,B_s^2\, \hat{\rho}^2\, M\,c^2
\end{equation}
where $c_{sv} = 0.340 \,\frac{m_q\,c}{\hbar} \, r_{bag}$ , and
$\hat{\rho} = \frac{\rho}{\rho_0}$. The coefficient 0.340 is a
dimensionless constant which can be derived from the MIT bag model
\cite{Guichon}. In that work, nuclear matter saturation and some
properties of finite nuclei are fitted with a bag radius of
$r_{bag} = 0.8 fm$.  This means that: $c_{sv} = 0.340 \cdot
\frac{313}{197} \cdot 0.8 = 0.432, $ essentially the same as 0.427
obtained using the ENJL model. Their value { for} the volume integral
of the scalar coupling is $\widetilde{V}_s= -1719\, MeV fm^3$,
very close to the ENJL result, $\widetilde{V}_s= -1705\, MeV
fm^3$.
\vspace{-0.25cm}
\subsection{Relativistic Mean Field Models for Equation of State}
\vspace{-0.25cm}
In the non-linear and derivative coupling mean field models, we also get a term quadratic in the density.
In the strong coupling limit, neglecting kinetic energy, we can write for the energy per particle:
\begin{equation}
W = M-1\,+ \,\frac{1}{2}\,B_v\,\hat{\rho}\,+\,\frac{(M-1)^2}{2\,B_s\,\hat{\rho}\,M^\alpha}
\end{equation}
The effective mass is obtained by setting $\frac{dW}{dM} = 0.$ \\
In the low density limit, the effective mass is given by:
\begin{equation}
M = 1\, - \, B_s\,\hat{\rho}\,+ \, \frac{3}{2}\, \alpha\, B_s^2\,\hat{\rho}^2\,+ \,...
\end{equation}
and
\begin{equation}
W = \frac{1}{2}\,(B_v\,-\,B_s)\,\hat{\rho}\,+\,\frac{1}{2}\,\alpha \, B_s^2\,\hat{\rho}^2\,+\,...
\end{equation}
Making a comparison with Eq. (\ref{saturating}), we obtain $c_{sv} = \frac{1}{2}\,\alpha $.
In the original derivative coupling model
\cite{Zimanyi90}, we get $\alpha = 2$, and $c_{sv} = 1$, which 
seems too large. However, in the hybrid model developed by
Glendenning {\it et al} \cite{Glendenning92}, we have $\alpha = 1$, which corresponds to
$c_{sv} = 0.5$. This is close to our phenomenological result. On the other hand, with the other EOS parametrizations, $c_{sv}$ has different values.

\vspace{-0.25cm}
\subsection{Connection of ECS model to other relativistic chiral models}
\vspace{-0.25cm}
In the present form the Lagrangian density (\ref{lagr2}) contains terms coupling the meson
fields to the fermion fields in a non-linear way. Using 
the equations of motion and
keeping only quartic terms in the meson fields we get,
\begin{eqnarray}
{\cal L}&=&\bar\psi(i\gamma^\mu\partial_\mu)\psi
-g_s\,(\sigma\,\bar \psi \psi+\bar\psi i\gamma_5\vec\pi\cdot\vec\tau\,\psi)-
g_v \, \bar\psi V^\mu\gamma_\mu\psi\,\nonumber\\
&+&{1\over2}(\partial^\mu\sigma\partial_\mu\sigma+
\partial^\mu\vec\pi\cdot\partial_\mu\vec\pi)
-{1\over2}m_s^2(\sigma^2+\vec\pi\cdot\vec\pi)-{1\over4}V^{\mu\nu}V_{\mu\nu} + \frac{1}{2}
m^2_v\, V^\mu V_\mu\,\nonumber\\
&+& {1\over2}
\frac{g_s\, m_v^4}{g_v^2} \, a_1
(\sigma^2+\vec\pi\cdot\vec\pi)\, V^\mu\, V_\mu.
\label{lagr1}
\end{eqnarray}
This Lagrangian density, except for the meson kinetic terms,  is
similar  to the one  introduced by Boguta \cite{boguta83} which,
by including a   scalar-vector coupling, was  able to reproduce
nuclear matter saturation properties. However, in \cite{boguta83}
an  effective mesonic self-interaction corresponding to  the
``Mexican hat''  is present.

In \cite{furn93} a generalization of the model proposed in \cite{boguta83}, including a
``bare'' vector mass, a quartic term in vector field and allowing for different vector-scalar and
vector-nucleon couplings, was studied.  It was argued that a linear realization of chiral
symmetry was too restrictive and, although nuclear matter properties may be reproduced,
there are problems with the  properties of finite nuclei.
{In our formulation, however, and with  our choice of parameters
we have shown that not only nuclear matter saturation properties
are reasonably explained,  but also surface properties.}

\vspace{-0.5cm}
\section{Conclusions}
\vspace{-0.5cm}
We have studied  an extension of the NJL model which yields reasonable saturation of nuclear
matter. An effective density dependent coupling constant is
obtained which pushes
chiral symmetry restoration to higher densities.
Two variants of the model defined by appropriate
sets of the parameters, $G_s,\,G_v,\,G_{sv},\,$ and $ \L,$ have been studied numerically.
For Model I, the values of the defining parameters, the vacuum properties, the properties at saturation are
given in Tables \ref{tab:exp} and \ref{tab:sur}.
{ An extension of the chiral sigma model has been developed which is based on the extended NJL model and
 fits not only the properties at saturation but also the empirical nuclear
surface energy and thickness. 
The term in $G_{sv},$ responsible for the density dependence of the effective
coupling constant, plays an important role in pushing to higher
densities the restoration of chiral symmetry and in lowering the incompressibility.

The relation
of the present model to the relativistic chiral model proposed by
Boguta \cite{boguta83} was presented. Finally we have discussed
asymmetric matter within the present model. 
 
 We have shown the connection of the ENJL
model with other relativistic models including QMC, ECS,
non-linear and derivative coupling mean field models. In order to
further test the model it is important to calculate other
properties of  finite nuclei and its performance at finite
               temperature. The critical temperature for chiral symmetry
               restoration at zero density is easily calculated to be
               $T_c=196$ MeV, an encouraging result.
Application of the model to
neutron-star matter requires the inclusion of strangeness and
beta-equilibrium.
\vspace{-0.5cm}
\section*{Acknowledgments}
\vspace{-0.5cm}
The authors would like to thank Mitja Rosina for fruitful discussions.
This work was partially supported by  FCT and FEDER under the
projects POCTI/FIS/451/94, POCTI/35308/FIS/2000, and POCTI/FP/FNU/50326/2003.

\end{document}